# A data-driven microscopic on-ramp model based on macroscopic network flows

Niklas Kolbe[a], Moritz Berghaus[b], Eszter Kalló[b], Michael Herty[a], Markus Oeser[b]

[a]RWTH Aachen University, Institute of Geometry and Practical Mathematics, Templergraben 55, 52062 Aachen, Germany
[b]RWTH Aachen University, Institute of Highway Engineering, Mies-van-der-Rohe-Strasse 1, 52074 Aachen, Germany

## Abstract

While macroscopic traffic flow models consider traffic as a fluid, microscopic traffic flow models describe the dynamics of individual vehicles. Capturing macroscopic traffic phenomena remains a challenge for microscopic models, especially in complex road sections such as on-ramps, In this paper, we propose a microscopic model for on-ramps derived from a macroscopic network flow model calibrated to real traffic data. The microscopic flow-based model requires additional assumptions regarding the acceleration and the merging behavior on the on-ramp to maintain consistency with the mean speeds, traffic flow and density predicted by the macroscopic model. To evaluate the model's performance, we conduct traffic simulations assessing speeds, accelerations, lane change positions, and risky behavior. Our results show that, although the proposed model may not fully capture all traffic phenomena of on-ramps accurately, it performs better than the Intelligent Driver Model (IDM) in most evaluated aspects. While the IDM is almost completely free of conflicts, the proposed model evokes a realistic amount and severity of conflicts and can therefore be used for safety analysis.

# 1   Introduction

Traffic flow modeling has a wide variety of applications such as transportation planning, traffic management and road safety analysis. Macroscopic models depict aggregated quantities like traffic flow, traffic density and mean speed (Aw & Rascle, 2000; Lighthill & Whitham, 1955; Richard, 1956; Zhang, 2002), whereas microscopic models describe the behavior of individual vehicles (Bando et al., 1995; Jiang et al., 2001; Treiber et al., 2000). While macroscopic models are more efficient to predict the traffic flow in a large road network, microscopic models are required for studying traffic flow in detail in specific traffic scenarios involving, for example, complex geometries. Macroscopic models make several simplifying assumptions, for example they neglect variability with respect to acceleration (Mohan & Ramadurai, 2013; Papageorgiou, 1998). Microscopic models describe the dynamics of vehicles in individual detail, but they do not necessarily reproduce empirical macroscopic traffic patterns such as the capacity drop (Saifuzzaman & Zheng, 2014). Under simplifying assumptions macroscopic traffic models can be derived as infinite particle limit of microscopic models, see e.g. (Colombo & Rossi, 2014). To combine the advantages of both models, it is promising to transform macroscopic models into microscopic models assuming that all vehicles have the same speed and spacing by converting traffic flow into time headways and traffic density into spacings.

While traffic modeling is very advanced on highway and motorway sections, complex geometries such as junctions lack the same degree of description because of the differing traffic behavior at these infrastructure elements. A macroscopic model for on-ramps was recently developed based on real data collected on German highway section in road work zones (Herty & Kolbe, 2022). In this paper we use this modeling approach to derive a new microscopic car-following model for on-ramps and we discuss the additional assumptions that are required on the microscopic scale.

The validation of traffic models usually happens on the macroscopic level and asks questions such as: does the model produce the expected traffic flow and density as real data? It is also necessary that a microscopic model describes driver behavior on the microscopic scale accurately. Different metrics are required to evaluate such microscopic models than the ones used for macroscopic models.

To use traffic flow models for road safety analysis, they must be able to predict the number of accidents or conflicts on a given road section or network (Mahmud et al., 2019). While in many microscopic models driver conflicts can be considered, they are mostly built to be accident free, which means vehicles will behave in a way that accidents cannot happen while simulating traffic (Calvert et al., 2020). With this simplification, the models may not be suitable for road safety analysis, for which models are sought that are capable to reflect risky driver behavior. Validating microscopic models can happen based on the amount of risky behavior produced. Risky behavior can be captured with so called surrogate safety measures (SSM), which quantify the risk of a collision between two interacting vehicles for example averaged over time or space (Mahmud et al., 2017). The closer the vehicles are to a collision, the less time they have to react, and when they react the more extreme maneuver such as strong braking or steering they would need to avoid a crash. The spatial distribution and severity of conflicts can be useful criteria to evaluate the model performance in terms of safety.

In this paper, we validate the proposed model in terms of speeds, accelerations, distribution of lane changes, and severity of conflicts and we compare its predictions with the Intelligent Driver Model (IDM). The rest of this paper is organized as follows: In the Methods section we first describe our data, the data-driven macroscopic junction modeling and our new model transferring these principles to microscopic models. We then give details on the implementation and the metrics we use for the model evaluation in the same section. In the Results and Discussion section we compare real traffic data at an on-ramp to simulation results obtained by an established microscopic model and our new flow-based approach.

## 2 Methods

### 2.1 Trajectory data

We use vehicle trajectory data from the German freeway A565 near the interchange Bonn-Beuel that was collected in May 2019 by drone photography (see **Figure 1**) (Lamberty et al., 2023). Vehicle types and trajectories were obtained from image processing after which the raw data points were interpolated by third order polynomials. A stretch of approximately 270 meters covering (including the entry lane) 4 unidirectional lanes was recorded. A total of 31 data sets were obtained from recordings over 4 days that took place either in the morning or in the afternoon, each one covering the traffic over periods of approximately 5 minutes. The number of entering vehicles varied with the time of day, on average 44 vehicles passed the junction per minute throughout the data.

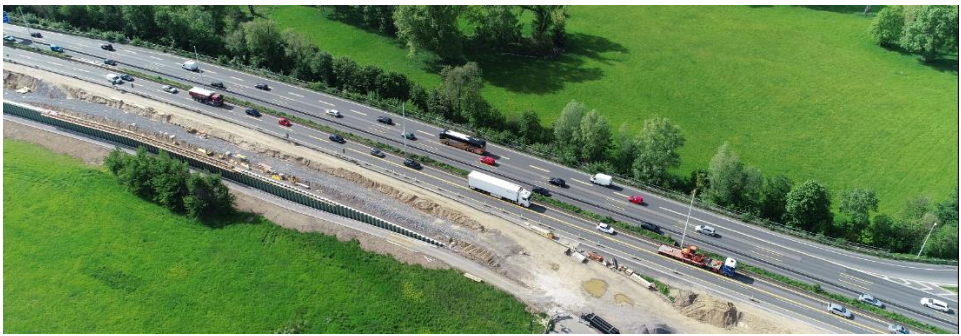

**Figure 1 Snapshot of the on-ramp (topmost lane, travel direction from right to left)**

### 2.2 Data-driven macroscopic junction model

The macroscopic model we consider is based on the approach in (Herty & Kolbe, 2022). In this work a junction model has been introduced that relies on vehicle trajectory data. As has been the case in this work, we focus here on the scenario of an on-ramp on a motorway, which can be classified into the following three segments: 1) the entry lane allowing vehicles to enter the motorway, 2) the stretch of motorway before vehicles from the entry lane can enter and 3) the stretch of motorway, at which vehicles from the on-ramp enter or have entered the freeway, respectively. Segments 2 and 3 amount to a short stretch, at which no further on-ramps or exits are located, corresponding to the previously described data set, in which these segments have a total length of approximately 300 meters.

In the macroscopic description no distinction between the individual lanes within the segments is made (in a realistic scenario, the on-ramp consists of a single lane, while both segments 2 and 3 cover at least two lanes). At each segment the traffic is governed by the Lighthill-Whitham-Richards (LWR) model

$$\rho_t + \left(\rho v_s(\rho)\right)_x = 0, \tag{1}$$

which describes the time and space (only a single space dimension is considered) dependent traffic density by means of a partial differential equation (Lighthill & Whitham, 1955). From the trajectory data vehicles are counted in fixed control volumes and their average velocities are computed, which allows to fit the parameters of the Greenshields fundamental diagram

$$v_s(\rho) = v_{max}^s \left(1 - \frac{\rho}{\rho_{max}^s}\right) \tag{2}$$

for each of the three segments. The model assumes that the three segments meet and therefore the traffic densities couple at a single fixed spatial point. The coupling is modeled by means of a Riemann solver,

$$\mathcal{RS}: (\rho_0^1, \rho_0^2, \rho_0^3) \mapsto (f_0^1, f_0^2, f_0^3) \tag{3}$$

mapping densities at the junction to coupling fluxes that govern the behavior at the junction. In numerical simulations the Riemann solver is applied to trace data at a given time t in order to compute boundary fluxes at the time $t + \Delta t$. A plethora of such coupling models has been evaluated; for simplicity we focus in this work on the flow maximization approach (Garavello et al., 2016), which maximizes the flux at the junction while assuring that both Kirchhoff conditions as well as the demand and supply conditions (Lebacque, 1996) hold. These conditions ensure mass conservation at the junction and the demand and supply conditions particularly depend on the fundamental diagrams corresponding to the three segments. The coupling model includes the parameter $\beta$, which is estimated from the traffic data and constitutes a priority rule in case of congestion in the outgoing segment. As the model has been originally adjusted to the same on-ramp data that is considered in this work it is used here with all estimated parameters as basis for a novel microscopic ramp model.

## 2.3 Transformation to microscopic model

Microscopic traffic flow models describe the dynamics of individual vehicles. Typically, the dynamics are divided into a longitudinal component, which is described by a car-following model, and a lateral component, which is described by a lane change model. Lateral movements within the lane are usually neglected. We consider a car-following model that determines the vehicle accelerations $a$ based on the longitudinal positions $x$, speeds $v$ and vehicle lengths $L$ of a vehicle pair (leader with index $L$ and follower with index $F$ on the same lane), i.e.,

$$a_F = f(x_L, x_F, v_L, v_F, L_L, L_F) \tag{4}$$

where the function $f$ is specified below. Together with the kinematic relationships $a_F = \dot{v}_F$ and $v_F = \dot{x}_F$, a car-following model is a system of ordinary differential equations (ODEs). Discretizing these relations in time, the model can be stated using the following update formulas for speed and position of a follower:

$$v_F(t + \Delta t) = v_F(t) + a_F(t) \cdot \Delta t \tag{5}$$

$$x_F(t + \Delta t) = x_F(t) + \frac{(v_F(t) + v_F(t + \Delta t))}{2} \cdot \Delta t \tag{6}$$

In the following we describe how information from a macroscopic model can be included in a microscopic model. Therefore, we make use of the following relations: the flow is the inverse of the mean (gross) time headway ($Q = 1/T$) at a specific location, i.e., the difference between the times two subsequent vehicles reach a fixed reference point. The density is the inverse of the mean (gross) spacing ($D = 1/S$) at a specific time, where the spacing S refers to the difference between the positions of two subsequent vehicles. The macroscopic fundamental equation of traffic flow ($Q = v \cdot D$) then corresponds to the microscopic equation:

$$v = \frac{S}{T} \tag{7}$$

This equation is valid for the mean values of speed, spacing and time headway as well as for the speed, spacing and time headway of individual vehicles.

We use the equation to model the speed of microscopic vehicles within a predefined stretch of road at the time $t + \Delta t$ from the spacing at the microscopic level and the at time $t$ as predicted by the macroscopic model:

$$v_{F,Model}(t + \Delta t) = S(t) \cdot Q(t) = \left(x_L(t) - x_F(t) - \frac{L_L + L_F}{2}\right) \cdot Q(t) \tag{8}$$

Here $L_L$ and $L_F$ denote the lengths of the leading and the following vehicles, respectively. The flow as a macroscopic quantity is the same for all vehicles at the considered stretch of road but the spacing may be different depending on the conditions at time $t$. As a result, the speed of each vehicle may also be different according to **Equation 7**. Nevertheless, the mean values of speed, spacing and time headway fulfil **Equation 7**.

In order to avoid sudden speed changes, we model the acceleration using the Full Velocity Difference Model (FVDM) by Jiang et al. (Jiang et al., 2001). Acceleration consists of two components. The first term states that vehicles accelerate proportionally to the difference between their current speed and the target speed. The second term reduces the acceleration proportionally to the speed difference between leader and follower in order to avoid collisions if the leader is very slow.

$$a_F(t) = \frac{v_{F,Model}(t) - v_F(t)}{\tau} - \lambda \cdot (v_F(t) - v_L(t)) \tag{9}$$

In this formula $\tau$ is the adaptation time that determines the sensitivity to speed differences between the current and the modelled speed and $\lambda$ is the sensitivity to speed differences between leader and follower. This microscopic car-following model allows individually different driving behavior for each vehicle while fulfilling the macroscopic model.

For sections 2 and 3 (see above), where vehicles mainly follow the leader on the same lane, this model can directly be applied. For the on-ramp (section 1), however, there are two additional phenomena that require modifications to the model: (1) vehicles on the on-ramp adapt their speed and spacing to the vehicles on the adjacent through lane, and (2) vehicles on the through lane may reduce their speed and increase the gap when a vehicle on the on-ramp wants to merge. We model these two phenomena by virtually combining the on-ramp and the adjacent through lane to a single lane. That means the leader is the closest vehicle in front on either lane, and the flow is the sum of the flows on these two lanes. At the end of the on-ramp, the assumption that both lanes can be combined is valid because vehicles on the on-ramp merge to the through lane and therefore need to adapt their behavior to the vehicles on the through lane. At the start of the on-ramp, however, vehicles are still allowed to overtake and to maintain small gaps to vehicles on the adjacent lane. We incorporate this behavior by using a larger adaptation time at the start of the on-ramp. We define the adaptation time as the remaining time until the end of the on-ramp given the current speed:

$$\tau = (x_{EndOfOnramp} - x_F)/v_F \tag{10}$$

### 2.4 Model implementation

We use SUMO to build a traffic simulation of the previously recorded freeway on-ramp in Germany (see **Figure 2**). SUMO is an open-source traffic simulation software that can easily be extended and modified (Lopez et al., 2018). The road network is modelled using input data from OpenStreetMap. As a result, the simulated road has the same course, and the entry lane has the same length as in the real traffic scenario. The width of the lanes is consistently 3.0 m, only the leftmost lane is 2.6 m wide. The microscopic model is implemented using SUMO's traffic control interface (TraCI). For the evaluation of the proposed flow-based model's performance, we also simulate traffic using the IDM (Treiber et al., 2000) as car-following model combined with Erdmann's lane change model (Erdmann, 2014), which is the default lane change model in SUMO. The traffic demand is taken from a 15 min excerpt of the vehicle trajectory dataset of the real road section (Lamberty et al., 2023). or both the microscopic flow-based model and IDM, we use four vehicle types: Passenger cars, vans, trucks and buses. Our flow-based model employs vehicle positions and velocities that are extracted from SUMO at every time step (0.1 s). The speeds predicted by the models are then fed back into SUMO to adapt the speeds of the vehicles in sections 1 to 3. Outside of these sections (in front of and behind the on-ramp), vehicles behave according to the IDM.

Although we model the driving behavior at on-ramps, we do not explicitly model lane changes. Instead, the car-following model implicitly ensures that lane changes are possible by letting the vehicles on the entry lane adapt to the vehicles on the right through lane and vice versa. The lane changes are triggered by SUMO's default lane change model, which ensures that lane changes are only conducted when there is a sufficient gap on the target lane.

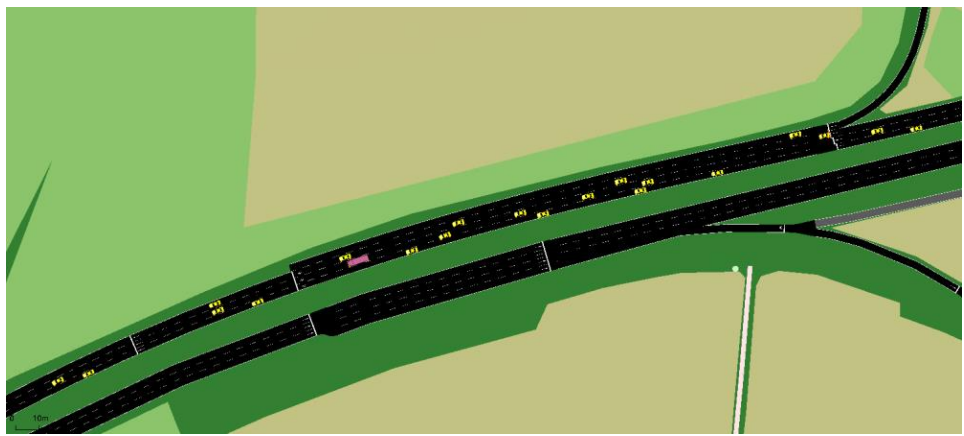

**Figure 2: Snapshot of the on-ramp in the SUMO simulation**

### 2.5 Model evaluation

We use Matlab for the model evaluation. For this purpose, we prepare the simulated traffic data in the same structure as the real traffic data. The simulation results contain information about each vehicles' category, size, location, speed and acceleration, the latter three being recorded every 0.2 seconds. The location information is given in SUMO's global coordinate system.

When vehicles on the main lane approach an on-ramp, they might adjust their behavior for a smoother flow, extending the available gap to help vehicles to merge onto the motorway. Therefore, we look at the speed and acceleration profiles, the lane changes and the safety of critical interactions in the different scenarios to judge the performance of the microscopic model to reproduce the real traffic data and in comparison to the Intelligent Driver Model. The location of the lane changes and the changes in speed and acceleration along road sections can indicate how good a traffic model simulates traffic on the microscopic level. For an insight, we create 5 m long sections on each lane to calculate the average local speed and average local acceleration values. Similarly, we create 20 m long sections where we count the number of lane changes.

On motorways, vehicles are mostly involved in rear-end conflict. In those cases, the application of surrogate safety measures to evaluate an interaction seems straightforward: There is a leader and a follower vehicle involved travelling behind each other in this explicit order. However, while lane changing, not every situation is clear, many times it is ambiguous which vehicle will take the 'leading' role. The determination of vehicle order or even the prediction of involvement of vehicles on the target lane gets even more difficult, when acceleration behavior is considered as well. Since vehicles accelerate on on-ramps and rather decelerate on main lanes for smoother merge, the aspect of this behavior cannot be neglected. To overcome the complexity of vehicle pair prediction in such situations, we defined vehicle pairs from the moment the position of the changing vehicle's left or right corner crossed the lane marking, even though its center point was still on the original lane. This predicts in which gap the changing vehicle is most likely going to be able to merge.

We use modified Time-to-Collision to evaluate vehicle interactions from the safety point of view. Time-to-Collision (TTC) is one of the most widely used surrogate safety measure, which calculates the nearness to a collision in seconds, based on the actual speed and initial net distance of following vehicles, suppose none of the vehicles would change their course (Hayward). The modified version of TTC, MTTC was developed by Ozbay et al. (Ozbay et al., 2008) and additionally takes into account the initial acceleration behaviour of both interacting vehicles, leading to a more realistic estimation. MTTC is calculated as follows:

$$MTTC = \begin{cases} \frac{D}{v_d}, if\ v_d > 0\ and\ a_d = 0 \\ \frac{-v_d \pm \sqrt{v_d^2 + 2 a_d D}}{a_d}, if\ a_d \neq 0 \end{cases} \quad (11)$$

where $v_d = v_{follower} - v_{leader}$ is the velocity difference, $a_d = a_{follower} - a_{leader}$ is the acceleration difference and D is the net distance between interacting vehicles at the moment of observation. When ad ≠ 0, the smallest positive value becomes the result of MTTC.

We calculate MTTC at every 0.2 seconds between vehicle pairs as described above. To see the development of critical interactions along the road, we calculate the average MTTC under 3 seconds with respect to predefined 20 m long sections. Therefore, we assign a leader-follower pair to the road section corresponding to the current position of the follower and take the minimal MTTC of that pair during the time the follower drives in that section. Then for each road segment the average over these values if under or equal to 3 seconds is computed.

# 3  Results and Discussion

To evaluate our proposed model, we compare real traffic data on the on-ramp (see Trajectory data) to simulations conducted with the proposed flow-based model and simulations with Intelligent Driver Model (IDM), see Model implementation. Our analyses focus on the spatial variations of traffic dynamics, and therefore consider temporal averages of key quantities with respect to the position on the on-ramp setting. To achieve this, we divide the road longitudinally into 5 m long sections.

First, we analyze the (longitudinal) velocities along the on-ramp. **Figure 3 (left)** shows that due to limited accuracy of the calibration both models overestimate the velocities of real traffic. The models, however, accurately represent the fact that the velocities on the on-ramp lane are smaller than on the through lanes. Comparing both models, we find that the variance in velocity between the lanes, which in the data ranges from 52.65 km/h on the on-ramp to 72.02 km/h on the second passing lane, is better represented by the new model predicting a range from 64.28 km/h to 82.66 km/h. Unlike the IDM model it assumes that the largest difference in average velocity is between the on-ramp and the main lane (from 64.28 km/h to 77.54 km/h), while in the data only a minor difference (52.65 km/h to 56.40 km/h) is observed. The largest difference in average velocities in the data is observed between the main lane and the first passing lane (from 56.40 km/h to 67.05 km/h), which is also the case in the simulations by the IDM model predicting a jump from 73.25 km/h to 81.36 km/h. These different model results might occur because the new model does not explicitly include differences between the through lanes.

In terms of acceleration, **Figure 3 (right)** shows that both models predict larger absolute accelerations compared to the data. This means that the observed behavior that drivers adapt their speed to the forthcoming on-ramp and therefore avoid large accelerations is not reflected in both models. In other words, both models tend to predict aggressive, reactive driver behavior. Since our model is derived from a macroscopic model, not explicitly accounting for acceleration, additional model assumptions in the future might achieve more realistic accelerations. Furthermore, we note that the model was only calibrated on macroscopic scale. Detailed microscopic calibration, which due to its large computational cost and complexity to quantify has not been realized in this analysis could be helpful to find a suitable balance between the two acceleration terms **Equation 9** and therefore obtain more realistic accelerations.

**Figure 3 (right)** also shows that in the results of our model the vehicles on the through lanes decelerate significantly in front of the on-ramp section to allow vehicles from the right to merge. This phenomenon can also be observed in real traffic data, but not as extreme. Moreover, the IDM model produces large negative accelerations at the end of the on-ramp lane because vehicles stop and wait for a suitable gap if they do not find one. Our proposed model does not show this unrealistic behavior.

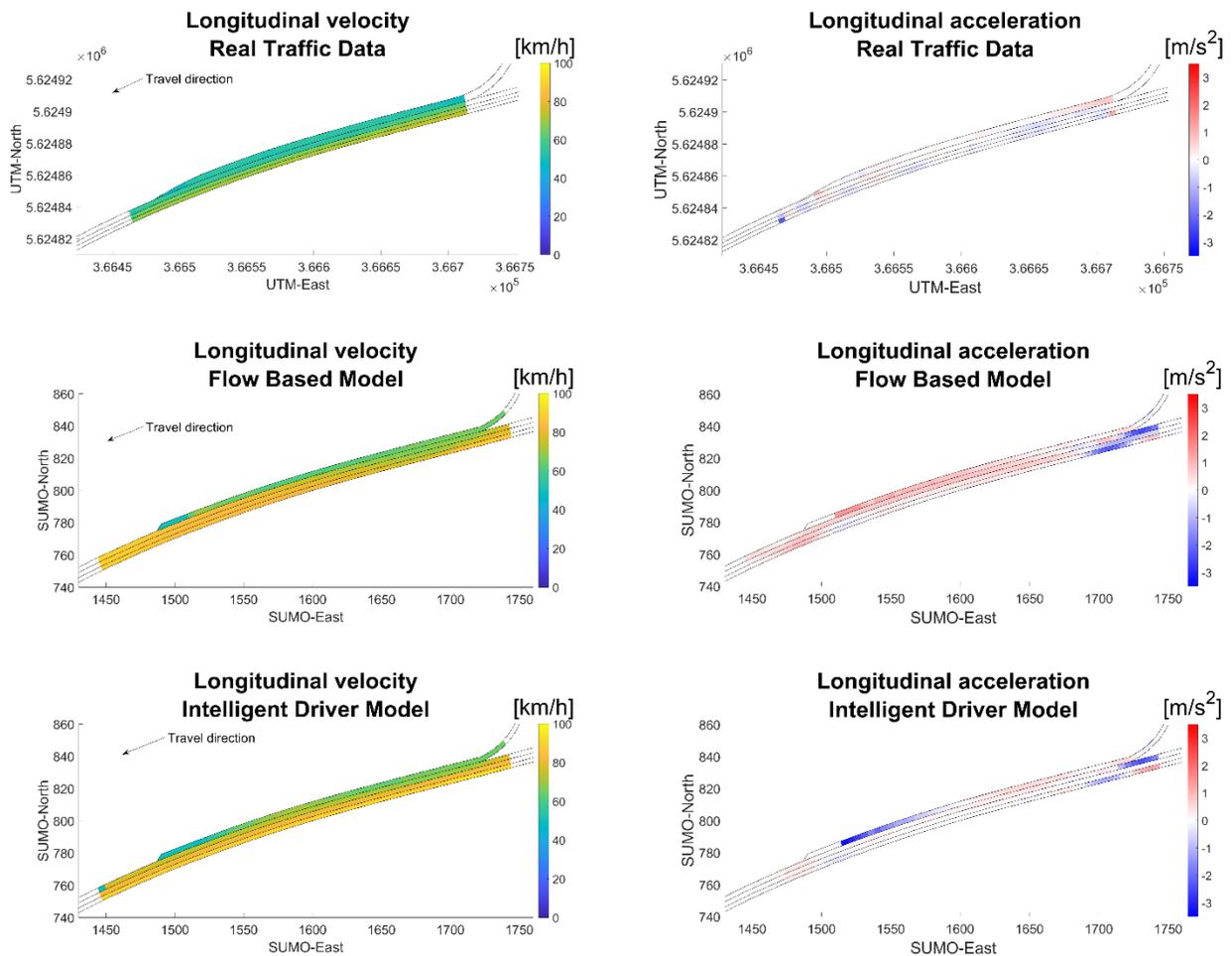

**Figure 3 Average longitudinal velocities (left) and accelerations (right) for real traffic data (top), our flow-based model (center), and IDM (bottom)**

In the real traffic data and the traffic simulated with the considered models no accident happens, which calls for other measures to evaluate traffic safety evaluation. With the help of Surrogate Safety Measures, areas with higher collision risk can be identified. **Figure 4 (left)** shows the average MTTC values under 3s within a given 20 m long section (see Model Evaluation section for details). If the average within these sections lies above 3 seconds, it indicates a safe, conflict-free area. In this case, the section is colored grey.

The real traffic data exhibits the lowest MTTCs and therefore the riskiest traffic scene. In traffic simulated with the new model, the on-ramp and main lane display the most dangerous area, but there is less to no disturbance on the two passing lanes. In traffic simulated with the IDM, only the first 20 m long sections on both the on-ramp and the main lane show low average MTTCs, while the rest of the analyzed stretch remains safe.

**Figure 5 (left)** shows the average MTTCs of the lanes: We calculated the average over 20 m long sections on the considered on-ramp scenario combining all lanes. Comparing the two models, we see that only three out of 15 sections have average MTTCs that are less than 3 seconds in case of the IDM model and thus generally safe driving behavior is assumed. The flow-based model predicts results that are more similar to the observed traffic scene in the data locally involving riskier driving behavior. Compared to real traffic the new model exhibits multiple locations at which a lower MTTC is pronounced. **Figure 5** (right) shows the total number of sections, that have average MTTC values between 0-1.5 seconds, 1.5-2 second, 2-2.5 seconds, 2.5-3 second and above 3 seconds. While the distribution of average MTTC achieved by the flow-based model differs from the one of the real traffic data, this is mainly because of the little number of conflicts on the passing lanes that the model predicts. **Figure 4 (right)** visualizes the changes from the entry lane to the main lane with respect to their location in the 20 m long sections on the considered on-ramp. In the simulations of the IDM model vehicles preferably change to the main lane as soon as possible. While a large part of the vehicles also changes very early in the simulation results of the flow-based model, more changes occur towards the end of the on-ramp, which is closer to the observed behavior in the real data.

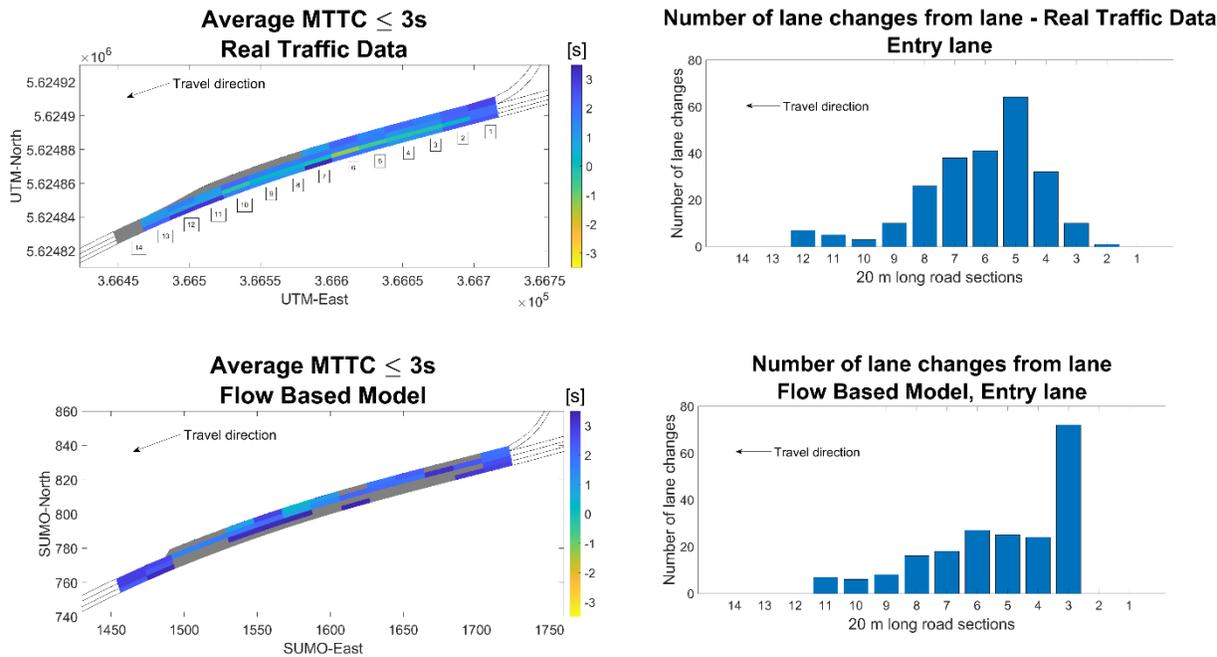

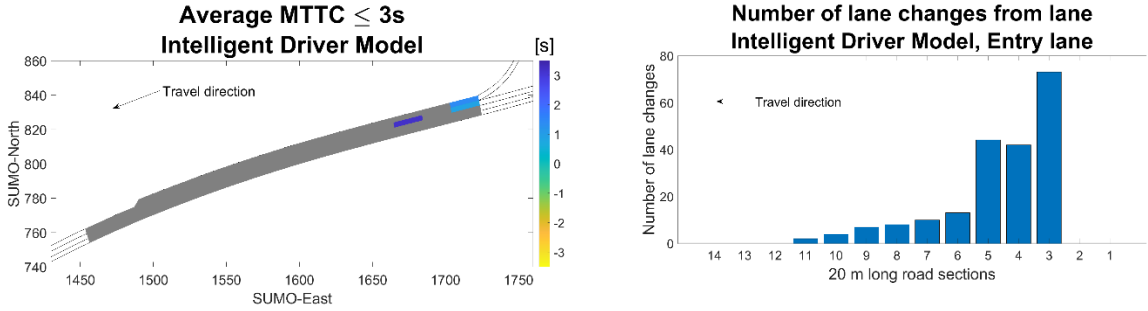

**Figure 4 (left) average MTTC values under or equal to 3s for real traffic data (top), our flow-based model (center), and IDM (bottom). Sections with no MTTC values under or equal to 3s are colored grey. (right) distribution of the positions of lane changes from the entry lane to the main lane for real traffic data (top), our flow-based model (center), and IDM (bottom)**

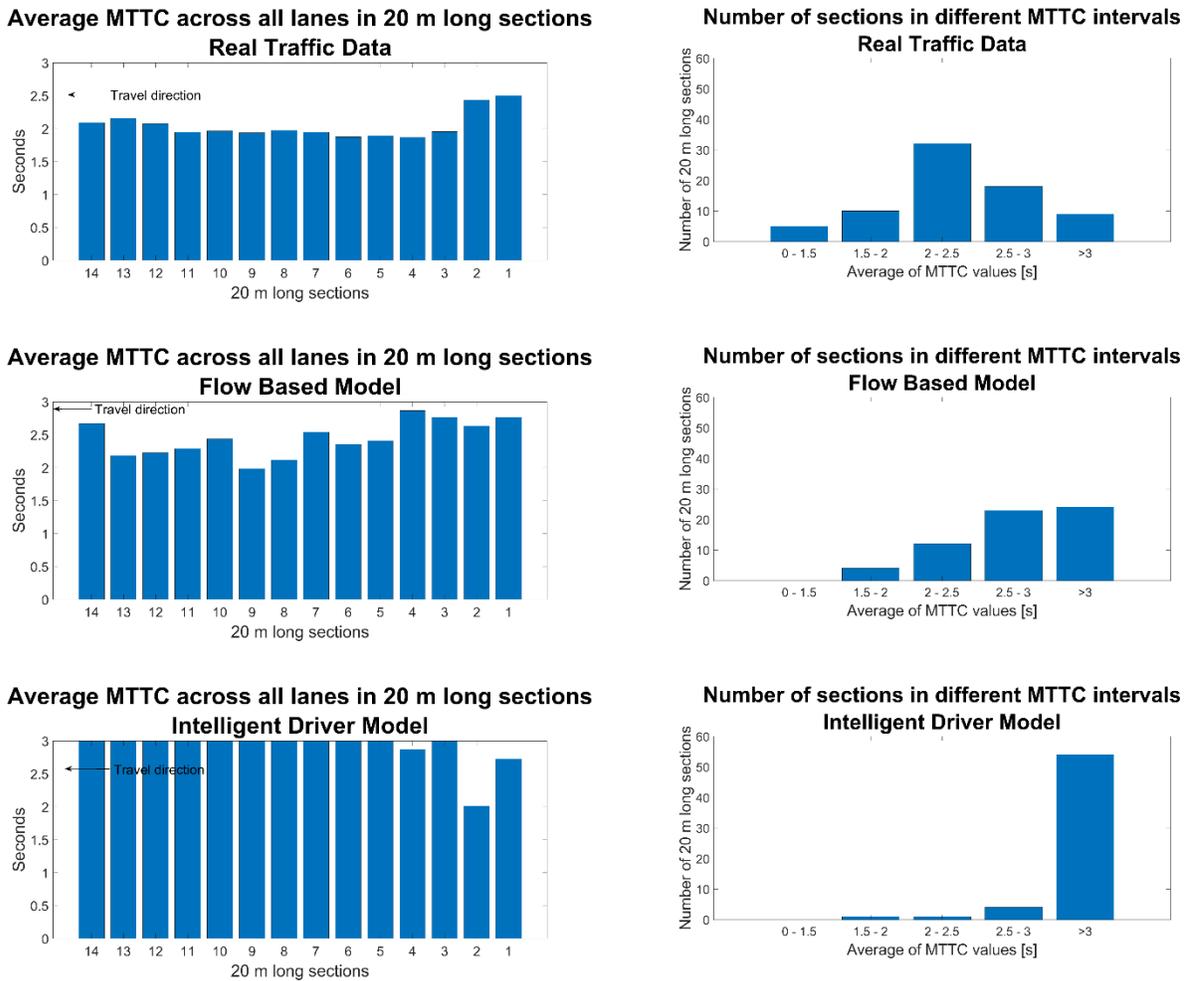

**Figure 5 (left) average MTTC values across all lanes under or equal to 3s for real traffic data (top), our flow-based model (center), and IDM (bottom). (right) count of 20 m long sections belonging to different MTTC intervals for real traffic data (top), our flow-based model (center), and IDM (bottom)**

# 4 Conclusions

In this paper, we proposed a microscopic car-following model for on-ramps derived from a macroscopic traffic flow model. The macroscopic model was calibrated based on real traffic data collected at a highway junction in Germany. The microscopic model requires only a few additional assumptions for aspects that are neglected in macroscopic models in order to ensure realistic merging behavior at on-ramps. By evaluating the proposed model in a traffic simulation, we found that the speeds and accelerations are larger than in real traffic, but more realistic than in the Intelligent Driver Model. The distribution of the lane change positions is also more realistic than in the IDM. The proposed model is suitable for road safety analysis as it creates a realistic amount and severity of conflicts. Since the calibration was performed on the macroscopic level, the accelerations are larger than in the real data. With microscopic calibration, a better resemblance between the model and real traffic might be achieved in future work.

Combining the advantages of both macroscopic and microscopic models is a promising approach to obtain a model that represents both microscopic driver behavior and macroscopic traffic phenomena well. However, even with thorough calibration, on-ramps remain a challenging road section to model.

## Acknowledgments

The work presented in this paper is part of the project NeMo (Neue Ansätze der Verkehrsmodellierung unter Berücksichtigung komplexer Geometrien und Daten - New traffic models considering complex geometries and data), project number 461365406, funded by the German Research Foundation (DFG). NK and MH further thank the DFG for the financial support through 320021702/GRK2326, 333849990/IRTG-2379, B04, B05 and B06 of 442047500/SFB1481, HE5386/18-1,19-2,22-1,23-1,25-1, ERS SFDdM035 and under Germany's Excellence Strategy EXC-2023 Internet of Production 390621612 and under the Excellence Strategy of the Federal Government and the Länder.

## Author Contributions

The authors confirm contribution to the paper as follows: study conception and design: M. Oeser, M. Herty; data collection: E. Kalló, M. Berghaus, N. Kolbe; analysis and interpretation of results: E. Kalló, M. Berghaus, N. Kolbe; manuscript preparation: E. Kalló, M. Berghaus, N. Kolbe. All authors reviewed the results and approved the final version of the manuscript.